\documentclass[a4paper]{jpconf}
\usepackage{graphicx}
\usepackage{amssymb}
\usepackage{amsmath}
\usepackage{bbm}

\newcommand{\mk}{\sc Mathematica}
\newcommand{\fr}{\sc FeynRules}

\newcommand{\bea}{\begin{eqnarray*}}
\newcommand{\eea}{\end{eqnarray*}}

\newcommand{\bpm}{\begin{pmatrix}}
\newcommand{\epm}{\end{pmatrix}}

\def\bsp#1\esp{\begin{split}#1\end{split}}
\def\be{\begin{equation}}
\def\ee{\end{equation}}
\def\del{\partial}

\newcommand{\psibar}{{\bar\psi}}

\usepackage{color}
\usepackage{ulem}

\begin{document}
\title{New developments in {\sc FeynRules}.}

\author{\underline{Adam Allou}l$^{1,2}$, Neil D Christensen$^3$, C\'eline Degrande$^4$, Claude Duhr$^{5,6}$, Benjamin Fuks$^{1,7}$}

\address{$^1$ Institut Pluridisciplinaire Hubert Curien/D\'epartement Recherches Subatomiques, Universit\'e de Strasbourg/CNRS-IN2P3, 23 Rue du Loess, F-67037 Strasbourg, France}
\address{$^2$ Groupe de Recherche de Physique des Hautes \'Energies (GRPHE), Universit\'e de Haute-Alsace, IUT Colmar, 34 rue du Grillenbreit BP 50568, 68008 Colmar Cedex, France}
\address{$^3$ PITTsburgh Particle physics, Astrophysics and Cosmology Center (PITT PACC), University of Pittsburgh, Pittsburgh, PA 15260 USA}
\address{$^4$ Department of Physics, University of Illinois at Urbana-Champaign,1110 West green Street, Urbana, IL 61801, United States}
\address{$^5$ Institute for Theoretical Physics, ETH Z\"urich, 8093 Z\"urich, Switzerland}
\address{$^6$ Institute for Particle Physics Phenomenology, University of Durham, Durham, DH1 3LE, United Kingdom}
\address{$^7$ Theory Division, Physics Department, CERN, CH-1211 Geneva 23, Switzerland }

\ead{adam.alloul@iphc.cnrs.fr, neilc@pitt.edu, cdegrand@illinois.edu, duhrc@itp.phys.ethz.ch, benjamin.fuks@iphc.cnrs.fr, fuks@cern.ch}

\begin{abstract}
The program {\fr} is a {\mk} package developed to facilitate the implementation of new physics theories into high-energy physics tools. Starting from a minimal set of information such as the model gauge symmetries, its particle content, parameters and Lagrangian, {\fr} provides all necessary routines to extract automatically from the Lagrangian (that can also be computed semi-automatically for supersymmetric theories) the associated Feynman rules.
These can be further exported to several Monte Carlo event generators through dedicated interfaces, as well as translated into a {\sc Python} library, under the so-called UFO model format,
agnostic of the model complexity, especially in terms
of Lorentz and/or color structures appearing in the vertices or of number of external legs.
In this work, we briefly report on
the most recent new features that have been added to {\fr}, including full support for spin-$\frac32$ fermions, a new module allowing for the automated diagonalization of the particle spectrum and a new set of routines
dedicated to decay width calculations.
\end{abstract}

\section{Introduction}
\vspace*{-17cm} \noindent 
\small{CERN-PH-TH/2013-235}
\vspace*{169mm}

The high-energy particle physics community is living nowadays in one of the key moments of its history.
The analysis of data acquired by the most powerful particle collider ever built, the Large Hadron Collider
at CERN,
is on-going and exciting results have already been established. For instance, the discovery
of a scalar particle with a mass of about 125~GeV featuring the
properties expected from the Standard Model (SM) Higgs boson was one of the major announcements in
2012~\cite{:2012gk,:2012gu}.
In parallel, a major effort is put into the exploration of possible beyond the Standard Model (BSM) signatures
hidden in data by comparing data to theoretical predictions.

In this context, Monte Carlo event generators play an important role as they allow to simulate
the behavior of both the Standard Model background and the new physics signals. While
these tools are ready to be used to simulate any SM process, the situation is however different
for BSMs where these models are not yet implemented in the Monte Carlo event generators.
The
{\fr} package~\cite{Christensen:2008py,Christensen:2009jx,Christensen:2010wz,Duhr:2011se,Fuks:2012im,
Alloul:2013fw,Christensen:2013aua} has been created with the aim of facilitating
this task. Starting by providing basic model information such as its underlying gauge symmetries,
field content, parameters and Lagrangian (that can also be semi-automatically derived
for supersymmetric theories), {\fr} then extracts all the interaction vertices from the model
Lagrangian
and exports the associated Feynman rules, through dedicated interfaces, to various high-energy
physics tools. Such interfaces exist for the {\sc CalcHep}~\cite{Pukhov:2004ca,Belyaev:2012qa},
{\sc FeynArts}~\cite{Hahn:2000kx}, {\sc MadGraph}~\cite{Stelzer:1994ta,Maltoni:2002qb,Alwall:2007st,%
Alwall:2008pm,Alwall:2011uj,deAquino:2011ub}, {\sc Sherpa}~\cite{Gleisberg:2003xi,Gleisberg:2008ta}
and {\sc Whizard}~\cite{Moretti:2001zz,Kilian:2007gr} programs.
Moreover, the model can also be translated to a {\sc Python} library under the Universal FeynRules Output (UFO)
format~\cite{Degrande:2011ua}. The UFO has been developed to overcome all
restrictions imposed by previously designed model formats for Monte Carlo programs.
It indeed allows for arbitrary color and/or Lorentz structures in the interaction vertices and any number
of external particles.

In this paper, we briefly review some of the new functionalities that will be included
in the new version of the package, {\sc FeynRules} 2.0, that
can already be downloaded from the website \texttt{http://feynrules.irmp.ucl.ac.be}.
We first describe the {\fr} program
in section~\ref{sec:fr}. Next, we detail the support for spin-$\frac32$ fields in
section~\ref{sec:spin_3_2_rarita_schwinger_fields} and a new module allowing for automated
spectrum generation in section~\ref{sec:automated_mass_diagonalization}.
Section \ref{sec:automatic_calculation_of_decay_widths} is dedicated to automated
decay width calculations and we finally summarize our work in
section \ref{sec:conclusion}.

\section{The {\fr} package}\label{sec:fr}
The {\fr} program is a {\sc Mathematica} package allowing for extracting
Feynman rules from a quantum field theory Lagrangian.
Additionaly to the improvements described in this paper, the latest version of {\fr} comes with a set
of tools directly connecting the model definition in {\fr} to a series of Monte Carlo programs.
Furthermore, it is also possible to store the model information in the form of a {\sc Python} library, under
the so-called Universal {\fr} output (UFO) format. This takes advantage of the object-oriented
nature of the {\sc Python} language and its ability to be interfaced with any other program.

Proper {\fr} model files follow a syntax inspired by the {\sc FeynArts} model format and encompass
information about the model gauge symmetries, field content and parameters.
From the input of the Lagrangian describing the dynamics of the model particles,
possibly given in a very compact form by means of quantities such as gauge covariant
derivatives, (super)field strengh tensors or even user-defined functions,
the {\fr} package is able to extract all the Feynman rules associated with the underlying
interaction vertices. The latter can then be exported, together with all the relevant model
information to several Monte Carlo and symbolic tools via dedicated interfaces.
In addition, supersymmetric models can be naturally and easily
implemented in a superspace module shipped with the code allowing to
declare chiral and vector superfields as any other field and containing a set of methods
dedicated to superspace computations.

All model publicly available can be found in the database located in the {\fr} website.

\section{Spin-3/2 Rarita-Schwinger fields} 
\label{sec:spin_3_2_rarita_schwinger_fields}
It is of critical importance for phenomenological purposes to have tools able to handle any new physics models,
no matter the structure of the interactions or the particle content. In particular, an automated
treatment, starting at the level of {\fr}, for spin-$\frac32$ particles was still missing until
very recently. The version 2.0 of {\fr} fills
this gap. The user can now use two new particle classes
\texttt{RW} and \texttt{R} to declare two- and four-component spin-$\frac32$ fields,
respectively. In parallel, the UFO conventions have been updated so that they account for
the arbitrariness associated with spin-$\frac32$ propagators, and recent versions of
both Monte Carlo event generators
{\sc CalcHep} and {\sc MadGraph}~5 support spin-$\frac32$ fields and their
interactions~\cite{Christensen:2013aua}.

The superspace module of {\fr}~\cite{Duhr:2011se}
has also been upgraded, and computations related to goldstino
and gravitino fields are now simplified within the {\fr} framework.
In the case of local supersymmetry, the interactions of the goldstino, a massless fermionic field
arising from supersymmetry breaking and those of the gravitino, a spin-$\frac32$ field predicted by
supergravity theories, are derived from the knowledge of the supercurrent,
a Majorana fermionic quantity $(J^\mu, \bar{J}^\mu)$ defined by
$$ J^\mu_\alpha = \sqrt2 D_\nu \phi^\dagger_i (\sigma^\nu \bar{\sigma^\mu}\psi^i)_\alpha + \sqrt2 (\sigma^\mu \bar{\psi}_i)_\alpha W^{*i} + g \phi^\dagger T_a \phi(\sigma^\mu \bar{\lambda}^a)_\alpha - \frac{i}{2}(\sigma^\rho \bar{\sigma}^\nu \sigma^\nu \bar{\lambda}_a)_\alpha v_{\nu \rho}^a\ . $$
In this expression, a sum over the model particle content is understood, and 
$v_{\nu \rho}$ denotes the field strength tensor for the vector boson $v_\mu$,
$D_\nu$ stands for the gauge-covariant derivative, $\sigma^\mu$ is the four-vector built upon the Pauli
matrices and $T_a$ are generators of the model gauge group taken in the appropriate representation.
Moreover, we generically denote scalar fields by $\phi$, gauginos by $\lambda$, the Weyl components
of the chiral supermultiplets by $\psi$ and $W^{*i}$ is the derivative of the hermitian conjugate of the
superpotential $W$. The interactions of the gravitino $\tilde G$ and the goldstino $\chi$ fields
are then given by
\be\bsp
  {\cal L}_{\chi} =&\ \frac{i}{2} \Big(\chi\sigma^\mu\partial_\mu\bar\chi - 
   \partial_\mu\chi \sigma^\mu\bar\chi\Big)
  + \frac{1}{2 v_1} \Big[ \chi \!\cdot\! \partial_\mu\tilde J^\mu 
  + \psibar_1 \!\cdot\! \partial_\mu\bar{\tilde J}^\mu \Big] \ ,\\
  {\cal L}_G =&\ \frac12 \epsilon^{\mu\nu\rho\sigma} 
    (\psi_{\tilde{G}\mu} \sigma_\rho \del_\sigma\psibar_{\tilde{G}\nu} +
     \del_\sigma \psi_{\tilde{G}\nu} \sigma_\rho \psibar_{\tilde{G}\mu})
  + {\cal N} \Big[ \psi_{\tilde{G}\mu} \!\cdot\! J^\mu 
  + \psibar_{\tilde{G}\mu} \!\cdot\! \bar J^\mu \Big] \ ,
\nonumber \esp \ee
that can be implemented in {\fr} following standard techniques.

The supercurrent $J^\mu_\alpha$ can be derived in {\fr} by calling,
in a {\sc Mathematica} session, the function \begin{verbatim}
 SuperCurrent[lv,lc,lw,sp,mu]
\end{verbatim} which takes as arguments the Lagrangians \texttt{lv} associated with the gauge sector of the
theory, \texttt{lc} associated with its chiral sector and
the superpotential Lagrangian \texttt{lw}. Moreover, the supercurrent spin and Lorentz
indices are given by \texttt{sp} and \texttt{mu}.
As it was shown in Ref.~\cite{Duhr:2011se}, the different pieces of the supersymmetric Lagrangian
can be straightforwardly obtained by means of standard methods of the superspace module of {\fr}.


\section{Automated spectrum generation} 
\label{sec:automated_mass_diagonalization}
When implementing in a given Monte Carlo tool a specific model, it is important to pay
special attention to the numerical values of the various model parameters. This usually
requires to compute and diagonalize mass matrices associated with the different gauge-eigenstates
of the model, which often mix. In practice, such calculations cannot be solved analytically
and it is necessary to rely on a numerical code. This issue has implied the
development of a new module of {\fr}
allowing, on the one hand, for the analytical extraction of all the model mass matrices
from the Lagrangian, and, on the other hand, to generate a self-contained {\sc C++} package,
dubbed {\sc ASperGe}\footnote{The acronym {\sc ASperGe} stands for {\bf A}utomated {\bf Spe}ct{\bf r}um
{\bf Ge}nerator.}, diagonalizing the mass spectrum and outputting both masses
and mixing matrices under a Les Houches compliant format~\cite{Alloul:2013fw}.

In order to employ these new functionalities, it is necessary to reorganize the information contained
in the model file. As an example, let us consider the mixing of the neutral SM gauge bosons
$B_\mu$ and $W_\mu^3$ to the photon $A_\mu$ and the $Z$-boson $Z_\mu$ states,
\bea  \begin{pmatrix} A_\mu \\Z_\mu \end{pmatrix}  = U_w \bpm B_\mu \\ W_\mu^3 \epm \ , \eea
which involves the mixing matrix $U_w$.
Initially, these relations could have been implemented in the definition of the fields themselves, as in
\begin{verbatim}
	V[11] == { 
	  ClassName -> B,
	  Definitions -> { B[mu_] -> -sw Z[mu]+cw A[mu]}, ....};
	
	V[12] == { 
	  ClassName -> Wi,
	  Definitions -> { Wi[mu_,1] -> (Wbar[mu]+W[mu])/Sqrt[2], 
	                   Wi[mu_,2] -> (Wbar[mu]-W[mu])/(I*Sqrt[2]), 
	                   Wi[mu_,3] -> cw Z[mu] + sw A[mu]}, ...},
\end{verbatim}
where \texttt{B} (\texttt{Wi}) stands for the gauge boson associated
with the $U(1)$ ($SU(2)_L$) subgroup of the SM gauge group,
\texttt{cw} and \texttt{sw} are the cosine and sine of the weak mixing angle, respectively,
and the dots stand for other options irrelevant with respect
to the mixings
(more details can be found in Ref.~\cite{Christensen:2008py}).
The new functionalities of {\fr} 2.0 offer a second way to treat particle mixings.
It relies on a variable \texttt{M\$MixingsDescription}
which consists of a list where each element is dedicated to a specific mixing relation.
Turning to our example, this list is given by
\begin{verbatim}
	M$MixingsDescription == { 
	  Mix["1u"] == {MassBasis -> {A, Z}, GaugeBasis -> {B, Wi[3]}, 
	                MixingMatrix -> UG, BlockName -> WEAKMIX}}
\end{verbatim}
We attribute a unique symbol (here \texttt{"1u"}) to each mixing relation
and define its properties through the options
\texttt{MassBasis} (refers to the mass eigenstates),
\texttt{GaugeBasis} (refers to the mixing gauge eigenstates),
\texttt{MixingMatrix} (defines the symbol to use for the mixing matrix which does not have to be
declared as a model parameter) and
\texttt{BlockName} (refers to the name of the Les Houches block containing the numerical
values of the elements of the mixing matrix that will be calculated by {\sc ASperGe}).
Additional options are also available and special cases are possible, depending on the spin of the fields
under consideration. We refer to Ref.~\cite{Alloul:2013fw} for their extensive description.

In addition to defining all mixing relations within the variable \texttt{M\$MixingsDescription},
a correct extraction of the model mass matrices implies to provide some knowledge about the fields
getting a
vacuum expectation value (vev) at the ground state of the theory.
This information is included in the variable \texttt{M\$vevs},
a list of two-component elements. The first of these components is the name of a
field and the second one is the symbol
referring to its vev. The model mass matrices can then be obtained by issuing, in the {\sc Mathematica}
session,
\begin{verbatim}
	ComputeMassMatrix[lagr]
\end{verbatim}
where \texttt{lagr} stands for the symbol of the model Lagrangian and
information associated
with a specific mixing relation (here \texttt{"1u"}) can further be accessed by typing
\begin{verbatim}
	MixingSummary["1u"]
\end{verbatim}

The source code for the {\sc ASperGe} package is generated via the command
\begin{verbatim}
	WriteASperGe[lagr]
\end{verbatim}
and consists in a set of model-dependent and model-independent files stored in a directory whose
name is made up from the model name and the suffix \verb?_MD?. The model-dependent files consist of
{\sc C++} commands allowing to define the various parameters of the model and the mass matrices
to be diagonalized. In contrast, the model-independent files contain all necessary routines
allowing {\sc ASperGe} to compute both the mass eigenvalues and the rotation matrices from the mass
matrices on the basis of the QR reduction algorithm as implemented in {\sc Gsl}.
In addition, these files also include methods to store these results in a Les Houches compliant
format. 

Finally, the numerical code can be compiled and executed by either typing \verb?RunASperGe[]? in the {\mk} session
or executing in a shell, after moving to the dicrectory where it is stored,
\begin{verbatim}
	make; ./ASperGe input/externals.dat output/output.dat
\end{verbatim}
The first of the two files (\texttt{input/externals.dat}) contains, under the Les Houches format, the model
free parameters together with their numerical values and the second one (\texttt{output/output.dat})
collects the results of the execution of {\sc ASperGe}.


\section{Automatic calculation of decay widths} 
\label{sec:automatic_calculation_of_decay_widths}
A careful design of a benchmark scenario for a given model requires, in addition
to providing a numerical value for the model external parameters and the calculation
of the mixing matrices necessary to diagonalize the mass spectrum,
the computation of the particle total widths. Furthermore, the latter must in general be given
together with the decay table of all
the unstable particles. To this aim, the version 2.0 of {\fr} includes a series of functions
dedicated to width calculations~\cite{duhr:2013xxxx}. The advantages of using {\fr} for
such a purpose are twofold. First, the new module calculates the analytical formulas associated
to all tree-level $1 \to 2 $ processes, so that the results are independent of any specific
benchmark point. Second, the calculations can be performed once for a given model and
reused every time the benchmark scenario is modified.

Assuming that the model Feynman rules have been stored and expanded over all flavor indices in a variable
denoted by \texttt{verts}, the associated set of tree-level $1\to 2$ decay widths 
are obtained by issuing
\begin{verbatim}      
	res = ComputeDecays[verts]
\end{verbatim}
The results (stored both in a variable \texttt{res}
and in the {\fr} variable \texttt{FR\$PartialWidths})
consist of a list of two-component elements. Their first
piece is another list of three fields \texttt{\{F1,F2,F3\}} indicating the decay process of interest
(\texttt{F1} $\to$ \texttt{F2 F3}) while their second piece is the analytical formula of the 
corresponding partial width. Each possible decay of a field into two other fields is considered
regardless of the numerical values of the particle masses.
The results thus include all possible cyclic permutations of \texttt{F1}, \texttt{F2} and \texttt{F3}.

A specific partial width or branching ratio
associated with the decay \texttt{F1} $\to$ \texttt{F2 F3}
can then be retrieved analytically by respectively issuing, in the {\sc Mathematica} session, one of the two
commands
\begin{verbatim} 
	PartialWidth[{F1, F2, F3}]
	BranchingRatio[{F1, F2, F3}]
\end{verbatim}
while the total width of a field \texttt{F1} can be computed by typing
\begin{verbatim} 
	TotWidth[F1]
\end{verbatim}
The corresponding numerical values are obtained by making use of the standard {\fr}
command \texttt{NumericalValue}. 
In order to update the numerical values of the width parameters from the content
of the variable \texttt{FR\$PartialWidths}, the user only has to call the
function \texttt{UpdateWidths[ ]}.

In addition, the decay information is by default also included in the
UFO library generated from {\fr}, this option being switched off by setting the option
\texttt{AddDecays} of the function \texttt{WriteUFO} to \texttt{False}. The generated UFO files include
a new file, \texttt{decays.py}, with all the calculated analytical formulas stored in the variable
\texttt{FR\$PartialWidths}.


\section{Conclusion} 
\label{sec:conclusion}
The constant and fast evolution in new physics model building activities has triggered the development
of programs such as the {\mk} package {\fr}, which aims to facilitate the development of BSM models
and their implementation within Monte Carlo tools. The soon-to-be-released 2.0 version of {\fr} contains
several important new features with respect to the previous release, in addition to the
already available superspace module, interfaces to the {\sc CalcHep}, {\sc FeynArts},
{\sc MadGraph}, {\sc Sherpa} and {\sc Whizard} programs and possibility to translate the full model
under the form of a {\sc Python} library, the so-called UFO format. Among the major new functionalities,
we have presented in this report a brief description of two modules, a first one dedicated to the automated
computation of decay widths and branching ratios and a second one addressing the extraction of the model
mass matrices and their diagonalization. In addition, we have also described some improvements related
to spin-$\frac32$ particles.


\section*{Acknowledgments}
We are grateful to the organizers of the ACAT2013 conference for arranging such a nice event.
This work has been partially supported by the Theory-LHC France initiative of the CNRS/IN2P3 and the
French ANR 12 JS05 002 01 BATS@LHC. A.A. has been supported by a Ph.D.~fellowship of
the French Ministry for Education and Research. Ce. D. was supported in part by the U. S. Department of Energy under Contract No. DE-FG02-13ER42001 and by the NSF grant  PHY-0757889; N.D.C. was supported in part by the LHC-TI under U.S. National Science Foundation, grant NSF-PHY-0705682, by PITT PACC, and by the U.S. Department of Energy under grant No. DE-FG02-95ER40896 and C.D. was supported by the ERC grant ``IterQCD''.

\section*{References}
\bibliographystyle{iopart-num.bst}
\bibliography{biblio}{}

\providecommand{\newblock}{}
\begin{thebibliography}{10}
\expandafter\ifx\csname url\endcsname\relax
  \def\url#1{{\tt #1}}\fi
\expandafter\ifx\csname urlprefix\endcsname\relax\def\urlprefix{URL }\fi
\providecommand{\eprint}[2][]{\url{#2}}

\bibitem{:2012gk}
Aad G {\em et~al.\/} (ATLAS Collaboration) 2012 {\em Phys.Lett.\/} {\bf B716}
  1--29 (\textit{Preprint} \eprint{1207.7214})

\bibitem{:2012gu}
Chatrchyan S {\em et~al.\/} (CMS Collaboration) 2012 {\em Phys.Lett.\/} {\bf
  B716} 30--61 (\textit{Preprint} \eprint{1207.7235})

\bibitem{Christensen:2008py}
Christensen N~D and Duhr C 2009 {\em Comput.Phys.Commun.\/} {\bf 180}
  1614--1641 (\textit{Preprint} \eprint{0806.4194})

\bibitem{Christensen:2009jx}
Christensen N~D, de~Aquino P, Degrande C, Duhr C, Fuks B {\em et~al.\/} 2011
  {\em Eur.Phys.J.\/} {\bf C71} 1541 (\textit{Preprint} \eprint{0906.2474})

\bibitem{Christensen:2010wz}
Christensen N~D, Duhr C, Fuks B, Reuter J and Speckner C 2012 {\em
  Eur.Phys.J.\/} {\bf C72} 1990 (\textit{Preprint} \eprint{1010.3251})

\bibitem{Duhr:2011se}
Duhr C and Fuks B 2011 {\em Comput.Phys.Commun.\/} {\bf 182} 2404--2426
  (\textit{Preprint} \eprint{1102.4191})

\bibitem{Fuks:2012im}
Fuks B 2012 {\em Int.J.Mod.Phys.\/} {\bf A27} 1230007 (\textit{Preprint}
  \eprint{1202.4769})

\bibitem{Alloul:2013fw}
Alloul A, D'Hondt J, De~Causmaecker K, Fuks B and Rausch~de Traubenberg M 2013
  {\em Eur.Phys.J.\/} {\bf C73} 2325 (\textit{Preprint} \eprint{1301.5932})

\bibitem{Christensen:2013aua}
Christensen N~D, de~Aquino P, Deutschmann N, Duhr C, Fuks B {\em et~al.\/} 2013
   (\textit{Preprint} \eprint{1308.1668})

\bibitem{Pukhov:2004ca}
Pukhov A 2004  (\textit{Preprint} \eprint{hep-ph/0412191})

\bibitem{Belyaev:2012qa}
Belyaev A, Christensen N~D and Pukhov A 2012  (\textit{Preprint}
  \eprint{1207.6082})

\bibitem{Hahn:2000kx}
Hahn T 2001 {\em Comput.Phys.Commun.\/} {\bf 140} 418--431 (\textit{Preprint}
  \eprint{hep-ph/0012260})

\bibitem{Stelzer:1994ta}
Stelzer T and Long W 1994 {\em Comput.Phys.Commun.\/} {\bf 81} 357--371
  (\textit{Preprint} \eprint{hep-ph/9401258})

\bibitem{Maltoni:2002qb}
Maltoni F and Stelzer T 2003 {\em JHEP\/} {\bf 0302} 027 (\textit{Preprint}
  \eprint{hep-ph/0208156})

\bibitem{Alwall:2007st}
Alwall J, Demin P, de~Visscher S, Frederix R, Herquet M {\em et~al.\/} 2007
  {\em JHEP\/} {\bf 0709} 028 (\textit{Preprint} \eprint{0706.2334})

\bibitem{Alwall:2008pm}
Alwall J, Artoisenet P, de~Visscher S, Duhr C, Frederix R {\em et~al.\/} 2009
  {\em AIP Conf.Proc.\/} {\bf 1078} 84--89 (\textit{Preprint}
  \eprint{0809.2410})

\bibitem{Alwall:2011uj}
Alwall J, Herquet M, Maltoni F, Mattelaer O and Stelzer T 2011 {\em JHEP\/}
  {\bf 1106} 128 (\textit{Preprint} \eprint{1106.0522})

\bibitem{deAquino:2011ub}
de~Aquino P, Link W, Maltoni F, Mattelaer O and Stelzer T 2012 {\em
  Comput.Phys.Commun.\/} {\bf 183} 2254--2263 (\textit{Preprint}
  \eprint{1108.2041})

\bibitem{Gleisberg:2003xi}
Gleisberg T, Hoeche S, Krauss F, Schalicke A, Schumann S {\em et~al.\/} 2004
  {\em JHEP\/} {\bf 0402} 056 (\textit{Preprint} \eprint{hep-ph/0311263})

\bibitem{Gleisberg:2008ta}
Gleisberg T, Hoeche S, Krauss F, Schonherr M, Schumann S {\em et~al.\/} 2009
  {\em JHEP\/} {\bf 0902} 007 (\textit{Preprint} \eprint{0811.4622})

\bibitem{Moretti:2001zz}
Moretti M, Ohl T and Reuter J 2001  (\textit{Preprint} \eprint{hep-ph/0102195})

\bibitem{Kilian:2007gr}
Kilian W, Ohl T and Reuter J 2011 {\em Eur.Phys.J.\/} {\bf C71} 1742
  (\textit{Preprint} \eprint{0708.4233})

\bibitem{Degrande:2011ua}
Degrande C, Duhr C, Fuks B, Grellscheid D, Mattelaer O {\em et~al.\/} 2012 {\em
  Comput.Phys.Commun.\/} {\bf 183} 1201--1214 (\textit{Preprint}
  \eprint{1108.2040})

\bibitem{duhr:2013xxxx}
Alwall J, Duhr C, Fuks B, Mattelaer O, \"Ozt\"urk G and Shen C~H {\em In
  preparation\/}

\end{thebibliography}
\end{document}